\documentclass[11pt]{article}

\usepackage[preprint]{acl}

\usepackage{times}
\usepackage{latexsym}
\usepackage[T1]{fontenc}
\usepackage[utf8]{inputenc}
\usepackage{microtype}
\usepackage{inconsolata}
\usepackage{graphicx}
\usepackage{booktabs}
\usepackage{array}
\graphicspath{{figures/}}

\newcommand{\pblock}{P\textsubscript{BLOCK}}
\newcommand{\plego}{P\textsubscript{LEGO}}

\title{Re-executing Coding-Agent Evaluations:\\Unchanged Decisions and Unresolved Policy Risk}

\author{Tianyu Ding \quad Yaocheng Tong \quad Muzhi Yang \\
\textbf{Fei Jiang \quad Lei Yang \quad Boya Yu} \\
Independent Researchers \\
{\normalsize\normalfont Correspondence: \href{mailto:tid16@pitt.edu}{\texttt{tid16@pitt.edu}}}}

\begin{document}
\maketitle

\begin{abstract}
Executable tests grade large language model (LLM) coding-agent patches for leaderboards, rewards and
self-improvement. Two preregistered re-execution studies returned negative results.
SWE-bench Pro v1's offline option rejected none of 59 accepted Go patches whose grading configuration changed. Their exploratory, unweighted
upper confidence limit is 4.95\% under candidate independence, leaving a 3\% exposed-patch risk unresolved;
the preregistered estimator uses inverse sampling-probability weights. An extension was inconclusive.
In a selected stratum, a reconstructed module cache built from the reference solution's dependencies rejected accepted alternatives from four models
that passed with their own dependencies. Re-executing 1,929 selected agent-version/task evaluations from the Darwin G\"odel Machine's
archive preserved best-agent selection and system-variant ranking, with search history and scores of evaluations not rerun fixed.
Exploratory host, image and adapted-grader comparisons also preserved them. Seven of 23 disagreements with archived verdicts
did not recur on a second host; rebuilt images changed 16 three-run majority verdicts (0.83\%) under four-CPU execution.
We provide an evidence-to-claim summary, lessons for grader audits and an anonymous replication package.
These results concern accepted patches and fixed-history decisions; general policy safety and adaptive-learning
robustness require broader evaluations.

\end{abstract}

\section{Introduction}
\label{sec:intro}

Executable tests decide whether a large language model (LLM) coding agent solved a task. Their verdicts feed leaderboards,
agentic reinforcement learning (RL;~\citealp{du2026legorl}) and self-improvement
selection~\citep{zhang2025dgm,wang2025hgm}. Verdicts depend on patches, grader images, caches, network policy,
reference test lists and execution resources. Network restrictions close routes agents used to retrieve
solutions~\citep{huang2026dsec}; packaging dependencies removes reliance on external network
state~\citep{du2026legorl}. Passing the reference (gold) patch validates only that solution: the same configuration
may reject another accepted patch. We test the consequences for recorded verdicts and decisions.

We preregistered two questions, sealing their protocols, code and inputs in pushed commits before
confirmatory data collection.
Study~1 replays officially accepted SWE-bench
Pro~v1 patches~\citep{deng2025swebenchpro} under its evaluator's offline option. It tests whether confirmed reversible losses reach 3\% on policy-valid tasks. Study~2 re-executes the
SWE-bench evaluations released with the Darwin G\"odel Machine (DGM;~\citealp{zhang2025dgm,dgmcode}). It tests whether rerun instability or central processing unit (CPU) contention changes the best recorded agent version or ranks its
three system variants (arms) differently. Both protocols predicted negative outcomes. Study~1 warned that a negative result would not establish
absence of an effect; Study~2 disclosed a separate project's pilot with no verdict flips on 194 evaluations of DGM's best agent version.

Study~1 observed no losses among 59 exposed Go patches on 25 tasks; the other 222 evaluable patches faced
identical configurations. The exposed-sample upper limit is 4.95\% under candidate independence; the extension was inconclusive.
A reconstructed gold-bound cache rejected 6 of 15 accepted patches in a selected stratum, all recovered with their own
allowed dependencies. Study~2's recorded endpoints remained unchanged; 23 archived verdicts differed from all six
original-host reruns, and seven of those disagreements did not recur on a second host.

We provide an evidence-to-claim summary and replication package. The selected cache cases
provide existence evidence on accepted patches for the dependency-coverage constraint documented in Lego-RL's gold-cache
tooling~\citep{legorlcode}.

\section{Background and hypotheses}
\label{sec:background}

\paragraph*{Executable tests as rewards}
Coding benchmarks and the RL environments built from them score a solution by applying the agent's patch to a repository
snapshot and running tests in a container: SWE-bench requires originally failing tests to pass (\texttt{FAIL\_TO\_PASS})
and originally passing tests to keep passing (\texttt{PASS\_TO\_PASS})~\citep{jimenez2024swebench}, SWE-rebench validates tasks in containers with an image per
task~\citep{badertdinov2025swerebench}, SWE-bench Pro targets harder, long-horizon
problems~\citep{deng2025swebenchpro}, Multi-SWE-bench adds languages~\citep{zan2025multiswebench}, and Lego-RL uses the
verdict as a terminal binary reward~\citep{du2026legorl}. The verdict comes from the tests \emph{and} from everything
around them: the image, the dependency resolver, the network, and the limits on CPU, memory and time.

\paragraph*{Hardened graders and their validation}
Two pressures push graders offline. Packages downloaded at grading time make results depend on mutable external
repositories~\citep{silva2024gitbugjava}, and open egress lets an agent fetch the upstream fix: DSec reports agents that
used module proxies to retrieve code from GitHub~\citep{huang2026dsec}. Lego-RL packages test dependencies into task
images~\citep{du2026legorl}, SWE-rebench V2 pre-builds task environments~\citep{badertdinov2026swerebenchv2}, the SWE-bench
Pro evaluator offers a no-network flag while the maintainers' latest protocol keeps network access because some Go tasks
fetch modules at test time~\citep{sweproos}, and SWE-Bench Pro Verified blocks code-hosting domains but keeps the
dependency services builds need~\citep{zheng2026sweproverified}. Such designs can be validated by checking that the
gold patch still passes and the empty patch (no agent changes) still fails. A gold-bound hermetic grader uses the gold patch's dependencies without network access;
an alternative needing another version can fail to build and receive a wrong fix's verdict.
Lego-RL's tooling states this limit for SWE-bench Pro tasks whose gold patch bumps a Go module: its cache is resolved
from the gold module manifest (\texttt{go.mod}), so ``a correct fix that bumps a module to a different version than gold still cannot
download it''~\citep{legorlcode}. In RL the same error assigns a failed rollout's reward to a solved one. Prior work models
false-negative rewards from imperfect verifiers~\citep{cai2025noisyrewards}; this error is structured, because it falls
on solutions whose dependencies differ from the gold patch's.

\paragraph*{Hypotheses and predictions}
Each preregistration fixes a decision rule, called a gate: a complete, valid result below its thresholds is a KILL (the
hypothesis is not supported), and a result that fails its protocol's completeness or validity conditions is
INCONCLUSIVE. \emph{H1} (Study~1): under SWE-bench Pro~v1's offline option, the weighted rate of confirmed reversible verdict losses
among evaluable accepted candidates on tasks where the policy is valid is at least 3\% (Section~\ref{sec:design}).
The initial protocol predicted a KILL (subjective probability about 0.85), revised before any confirmatory data to about
0.65 when the feasibility stage exposed a memory-limit risk, and predicted that most Go tasks whose gold patch bumps a
module would be ineligible. \emph{H2} (Study~2): rerun instability or
CPU contention alone reverses a recorded DGM decision, reproducibly (Section~\ref{sec:study2}). The preregistration predicted a KILL, with
instability and contention each on at most 1\% of cells. The extension of Study~1 (Section~\ref{sec:extension})
has no kill rule: it dropped the 3\% gate, its outcomes are ``no rejection observed'', ``rejection observed'' or
inconclusive, and it predicted that no rejection would be observed. Table~\ref{tab:overview} in Appendix~\ref{sec:overview} summarizes the two studies, the extension and the feasibility
stratum.

\section{Study 1 design: offline grading of SWE-bench Pro}
\label{sec:design}

\paragraph*{Units and terminology}
An \emph{accepted patch} passes the benchmark's tests (\emph{resolved}); \emph{unresolved} means those tests were not satisfied.
Evaluable patches reproduce acceptance with complete controls; exposed patches face different policy configurations.
An \emph{artefact} is an accepted candidate, gold/empty patch, or officially unresolved control.
Study~2's unit is an agent-version/task evaluation (Section~\ref{sec:study2}).

\paragraph*{Policies and arms}
A \emph{grader policy} fixes what the verifier container may reach. The task image (prepared container environment,
pinned by content digest), entry script, tests, parser, resolution rule and time limit stay fixed.
Official verdicts came from a networked grader that downloads missing Go modules. Our containers have no network,
so baseline arm A0 emulates that contract. For each artefact it seeds a read-only file
module proxy with the module-checksum entries in \texttt{go.sum} that the image's cache lacks, then adds, round by round, every file of each module
version the artefact's own verification requests. The bytes are fetched on the host from public module proxies and checked by the
Go toolchain against \texttt{go.sum}. A0 must reproduce the official verdict before a candidate counts. Arm
A1\textsubscript{BLOCK} (A1 in the confirmatory analysis), the restricted arm, applies the offline policy \pblock, which disables networking and changes nothing
else~\citep{sweproos}; it is the single confirmatory policy. Arm A2, the restored arm, provisions per candidate: it keeps the shipped
image and adds the candidate's own A0 mirror, excluding the task's own module path (the project under repair), whose later versions could contain the fix. (In the feasibility stage, A2 added the candidate's modules to the gold-bound cache, excluding the task's own module path.) For Go tasks whose gold patch edits \texttt{go.mod} or
\texttt{go.sum}, arm A1\textsubscript{LEGO} applies the gold-cache policy \plego, our reconstruction of Lego-RL's gold-bound module
cache~\citep{legorlcode}. It runs the builder's Go commands on the gold state offline, against a file mirror seeded from the gold
\texttt{go.sum} and filled request by request until nothing new is requested; the upstream builder runs them with network access. \plego{} is
analyzed descriptively. For non-Go tasks the emulation adds nothing, so A0, A1 and A2 are identical configurations and
the design observes only background disagreement there.
Figure~\ref{fig:arms} (Appendix~\ref{sec:overview}) shows the arms and repetitions.
\paragraph*{Validity of a policy for a task}
A restrictive policy can break a task for every solution. A task is \emph{eligible} under a policy only if the gold
patch resolves in all three repetitions under it and the empty patch fails in all three; only candidates on eligible
tasks count.

\paragraph*{Research questions}
The research questions are: RQ1 (validity), on which tasks does \pblock{} pass the gold patch? RQ2 (incidence): what fraction of accepted
candidates on eligible tasks does \pblock{} reject, reproducibly and with a network-unavailability signature in its restricted logs? RQ3 (gold-cache rejection): are
accepted alternatives rejected under \plego? RQ4 (repair): does A2 recover rejected candidates without newly accepting
officially unresolved controls?

\paragraph*{Corpus and sampling}
We use the public split of SWE-bench Pro~v1~\citep{deng2025swebenchpro} at a pinned revision, with the official run scripts
and parsers at a pinned commit~\citep{sweproos}. The maintainers publish the patches of nine agent runs with their
verdicts; a \emph{candidate} is a patch the official evaluation marked resolved, and we draw from the nine runs
(one had no available patch). The frame contains every task with a resolved candidate, minus pilot, feasibility and known-broken tasks; tasks
are ordered by a seeded hash, and the confirmatory sample is the first 120, reserved from metadata before this study
fetched any of their candidate patches. The preregistration discloses a pre-project screen that downloaded the Go
candidate diffs of eight runs and flagged manifest edits without executing anything; the draw does not use it. Each task contributes up to four candidates in a frozen round-robin order over models, its gold
and empty patches as controls, and up to one officially unresolved candidate as a negative control.

\paragraph*{Outcomes, estimator and gate}
Each arm runs three times, in a fresh container with four CPUs, 8 gibibytes (GiB) of memory, no network and the official
one-hour timeout, in a seeded order interleaved across arms. A Study~1 cell is an (artefact, arm) pair and is \emph{finished} if all three
repetitions completed the grader invocation, with no launcher error, timeout or out-of-memory kill and with parser
output.
A candidate is \emph{evaluable} if its task is eligible, its gold-baseline control and candidate A0, A1 and A2 cells
finished, and A0 resolved it three times. It is a \emph{confirmed reversible loss} if A1\textsubscript{BLOCK}
rejects it three times, A2 resolves it three times, and the restricted logs show the preregistered
network-unavailability signature (\texttt{NET\_RE}; Appendix~\ref{sec:sealing}). Candidates are weighted by their inverse within-task inclusion probability, estimated at draw time from availability
alone with 20,000 fixed-seed Monte Carlo draws; the primary quantity is the
weighted loss rate, with a one-sided 95\% lower bound from a task-cluster bootstrap. The gate requires at least 200
evaluable candidates from at least two models and passes only if the rate is at least 3\%, the lower bound exceeds 1\%,
and losses occur on at least five tasks. A complete, valid sample below these thresholds is a KILL. The study is
invalid, and so inconclusive, if the gold patch fails any baseline run on an eligible task, if more than 10\% of the
drawn tasks have an unfinished control, if more than 10\% have finished gold-baseline cells that do not reproduce the
gold verdict, or if its integrity audit does not accept the execution records as valid (PASS); a task on which only the restricted arm fails the gold patch is merely
ineligible. In the preregistered simulation, which assumed no attrition, the design would
have passed with probability 0.48 (independent losses) or 0.40 (clustered losses) at a true rate of 3\%, and 0.95 or
0.88 at 5\%.
The protocol, code, manifests, seeds, sample reservation, analysis script and predictions were sealed by a pushed
commit before any confirmatory data; Appendix~\ref{sec:sealing} gives the sealing, amendment and audit procedure.

\section{Study 1 results}
\label{sec:results}

The confirmatory run executed all 120 tasks of the reserved prefix: 664 artefacts and 5,997 container runs (93.99
container-hours), with no interrupted or deadline-truncated cell. Its integrity audit passed on the fourth round after
post-data appendices corrected reporting and wording, without changing runs.

\subsection{RQ1: Where the offline policy is valid}
Under \pblock, the gold patch passed all three restricted runs and the empty patch failed all three on 113 of the 120
tasks. Four tasks, all from one JavaScript repository, have incomplete control evidence because their test workers
exceeded the 8\,GiB limit. Of the three ineligible tasks with finished controls, two fail the gold patch in every arm,
the baseline included; the third, a Go task, passes it in all baseline and restored runs but fails every restricted run
with failed module downloads, so the offline option cannot grade its reference solution. Both invalidity shares stayed
below the preregistered 10\%: 4 of 120 tasks with an unfinished control (3.33\%) and 2 of 120 whose finished
gold-baseline cells do not reproduce the gold verdict (1.67\%).

\subsection{RQ2: No observed losses, limited exposure}
\begin{table}[t]
\centering
\footnotesize
\begin{tabular}{@{}lrr@{}}
\toprule
Quantity & Value & Threshold \\
\midrule
Evaluable candidates (of 312) & 281 & $\geq 200$ \\
Models among them & 7 & $\geq 2$ \\
Tasks contributing them & 110 & \\
\shortstack[l]{Verdict losses\\(0 of 3 restricted runs resolved)} & 0 & \\
Confirmed reversible losses & 0 & \\
Weighted loss rate & 0.00\% & $\geq 3\%$ \\
One-sided 95\% lower bound & 0.00\% & $> 1\%$ \\
Tasks with a confirmed loss & 0 & $\geq 5$ \\
\midrule
\shortstack[l]{11 baseline-unreproduced candidates\\treated as losses (sensitivity)} & 3.90\% & \\
\bottomrule
\end{tabular}
\caption{Study~1 confirmatory result (offline policy \pblock, 120-task prefix). Rates are over evaluable candidates; the
descriptive sensitivity row also counts the 11 unevaluable ones on eligible tasks.}
\label{tab:m0b}
\end{table}

Of 312 drawn candidates, 281 were evaluable, from seven models on 110 tasks (Table~\ref{tab:m0b}); 9 were on
ineligible tasks, 11 on tasks with incomplete controls, and 11 were not reproduced by the baseline arm. No evaluable candidate lost its verdict or varied across restricted or restored repetitions.
The complete, valid sample returns KILL: H1 is not supported. A preregistered sensitivity analysis that counts the
unresolved candidates on eligible tasks as losses gives a weighted 3.90\%; with eligibility and weights fixed, it covers only the 11 baseline-unreproduced candidates on eligible tasks.
It exceeds 3\% but does not establish the registered lower-bound and five-task conditions. The KILL therefore gives no safety assurance. The 31 unevaluable candidates may include policy-sensitive cases: baseline emulation can turn missing inputs into attrition.

\begin{figure}[t]
\centering
\includegraphics[width=\columnwidth]{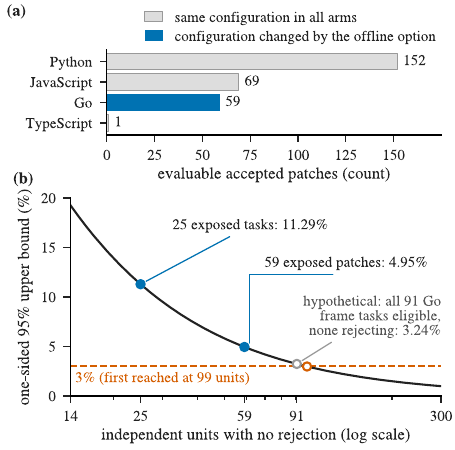}
\caption{Exposure and post-hoc bounds in Study~1. (a)~The 281 evaluable accepted patches by task language; only the 59
Go patches (25 tasks) faced a configuration that \pblock{} changes. (b)~One-sided 95\% Clopper-Pearson upper confidence limit
after zero rejections by the number of independent units. The two filled points are unweighted bounds for different
estimands: the patch rejection rate (59 patches) and the share of tasks with a rejection (25 tasks). The dashed line marks the 3\% of
the hypothesis as a reference level for both estimands, and the orange hollow marker is where the bound first reaches
it (99 units); the grey point is the hypothetical bound if all 91 Go tasks of the frame were eligible and none
rejected.}
\label{fig:study1}
\end{figure}

Only the 59 evaluable Go patches on 25 tasks faced different configurations (Figure~\ref{fig:study1}a);
all passed with the shipped image alone. The 222 evaluable Python, JavaScript and TypeScript patches faced identical
arms, with no background disagreement. Risk bounds for the exposed patches were not preregistered.
Post hoc and unweighted, the one-sided 95\% Clopper-Pearson upper confidence limit is 4.95\% on the rejection rate of these
candidates if they are independent, 11.29\% on the share of their 25 tasks with a rejection if tasks are independent,
and 27.17\% on the candidate rate if the 13 unevaluable Go draws count as rejections and candidates are independent.
Figure~\ref{fig:study1}b shows the two zero-event bounds. None excludes 3\%; 99 independent exposed candidates without a rejection would be needed. Under task independence,
the same frame cannot reach 3\%: even if all 91 Go tasks of the sampling frame were
eligible and none rejected, the
independent-task upper confidence limit would be 3.24\%. This hypothetical limit of that interval does not answer the registered
candidate-level question, and the design did not exercise package installation at grading time in other ecosystems.

\subsection{Extension to the remaining Go tasks}
\label{sec:extension}
A second preregistration, sealed before its data, replayed the 63 Go tasks that follow the prefix in the reservation
order with the same arms, counting \emph{any} failed restricted repetition of a reproduced candidate as a rejection.
The extension is \emph{inconclusive}: on one eligible task the gold patch failed one of its three baseline runs, which the
registered control rule does not allow, and we neither drop that task nor repeat the run. The registered tally is two
rejection tasks among 59 eligible, one of them that task; because the control failed, every bound is descriptive and the
predicted zero was not observed. Appendix~\ref{sec:extension-detail} gives the tally, the descriptive bounds and the
shared nil-pointer panic behind both failures.

\subsection{RQ3: Gold-bound caches}
The prefix contained one task whose gold patch bumps a Go module, with no loss under \plego; the extension added four,
the four on which the gold patch failed \pblock, and none of their nine evaluable candidates lost its verdict under the
reconstructed cache. The only cache-induced rejections come from the feasibility stage, which selected 11 such tasks on
purpose and is therefore no prevalence estimate. There, \pblock{} failed the gold patch on 8 of
the 11 tasks, while the reconstructed cache passed it on all 11. Of 15 baseline-reproduced accepted candidates on seven
of these tasks, six, on three tasks and from five runs of four models, failed all three runs under the cache and passed all three
once their own admitted modules were added. On one Flipt task, candidates from three models each added a version of
\texttt{Masterminds/semver} (v3.1.1 or v1.5.0) that is absent from the gold state's \texttt{go.sum}; each resolved three of three baseline
and restored runs and none under the cache. Appendix~\ref{sec:cache} describes the reconstruction and the parity checks it passed.

\subsection{RQ4: Candidate-aware provisioning}
Provisioning recovered all six feasibility-stage losses. In the confirmatory run there was nothing to recover, and the
preregistered safety criterion forbids any restored-arm acceptance when both other arms reject; it failed: one officially unresolved Python candidate passed one of three restored runs
after failing every other run. Its restored arm had no mirror and matched the other configurations, consistent with
background flakiness; the cause was not isolated and the frozen rule counts it. No negative control was observed consuming
provisioned modules (the official run scripts of 5 of the 112 tasks with a sampled negative control filter out the logged \texttt{go:~downloading} marker),
so that zero does not show such use to be absent. The labels have a limit of their own: 9 of the 112 sampled negative controls,
and 3 of the extension's 63, resolved in every run of every arm, so the replay does not reproduce the official
rejection for them; we did not classify the cause.

\section{Study 2: Re-executing a self-improvement archive}
\label{sec:study2}

A self-improving coding agent compares and promotes its versions using test verdicts.
Study~2 asks whether a released run's recorded decisions survive re-execution of frozen patches. A \emph{cell} is one archived
evaluation of an agent version on a task within a system arm; it can have multiple reruns. We separate three sources of disagreement: rerun instability, CPU contention, and drift from the archived grading environment.

\subsection{Design}
\paragraph*{Object and decisions}
DGM evaluates self-modified coding agents on SWE-bench Verified, up to 200 tasks~\citep{zhang2025dgm}.
Its archive~\citep{dgmcode} holds 245 agent versions (nodes), with 6,006 evaluations on 200 tasks in three arms:
the full system and ablations without open-ended exploration or self-improvement. The recorded arm bests are 100/200 for the full system, the 50.0\% that DGM reports, 23/59
without self-improvement and 14/60 without open-ended exploration. We reconstruct only decisions that the released
code defines and the archive records. DGM's score is the fraction of submitted tasks resolved. \emph{D1} (best-node selection) chooses the best of the
eight full-system nodes evaluated on the full task set (197 to 200 submitted tasks each; recorded 100/200 against a
runner-up of 87/200); \emph{D2} (arm ranking) ranks the three arms by the best score in each arm's final archive (the arm bests
above, each over its node's submitted tasks). A decision reverses if a challenger reaches or exceeds the incumbent.
The staged gate is not reconstructable: recorded history conflicts with the released threshold
(Appendix~\ref{sec:fidelity}).

\paragraph*{Replay}
We re-execute 1,929 recorded cells: every cell of the eight D1 nodes and of each arm's top three nodes by recorded score,
including ties at the cutoff, 15 nodes in all. Each cell runs three times with four CPUs (nominal) and three times with a quarter of a CPU (contention), each
run in a fresh network-isolated container with 8\,GiB. Two details affect replay fidelity, and both surfaced before the seal: DGM grades the patch after removing its diffs to test files, and the evaluation images (SWE-bench's September 2025 \texttt{v2} tag) commit setup edits the archived runs left uncommitted. Appendix~\ref{sec:fidelity} gives both checks and the required regrading agreement.

\paragraph*{Quantities and rule}
A majority verdict is the outcome of at least two of three repetitions. On each cell, Q1 is rerun instability
(non-unanimous nominal verdicts), Q2 is drift (the recorded verdict differs from the nominal majority), and Q3 is
contention disagreement (the contention majority differs from the nominal majority). We report counts of cells in each category.
Only Q1- and Q3-type reversals count toward the hypothesis. Starting from recorded verdicts, each Q1 comparison
uses the same repetition index across all unstable cells; Q3 substitutes contention majorities on cells without
drift. A reversal had to reproduce in a confirmation batch (Appendix~\ref{sec:fidelity}). Stable drift is reported
separately; no drift cell was also unstable. Nodes not re-executed keep their recorded scores.
The protocol, cell sets and code were sealed before any decision cell ran, after two rounds of design and code review.
We saw the earlier pilot's written assessment, not its data.

\subsection{Results}
\paragraph*{Evaluator and replay}
Regrading reproduced 5,999 of the 6,006 archived verdicts (99.88\%); all seven disagreements are on one Astropy task
whose reference test list was revised in the SWE-bench Verified copy used for regrading (Appendix~\ref{sec:versions}). All 11,574 runs completed
(68.55 container-hours) without timeouts or memory kills.

\paragraph*{Decisions unchanged}
Two cells showed rerun instability, and the same two are the only ones where the contention majority differs from the
nominal majority; both belong to one Sphinx task (sphinx-10466). Under independence
the one-sided 95\% upper bound is 0.33\% per cell for each quantity, and 2.35\% per task (1 of 200). Neither kind of
instability reversed D1 or D2, so no confirmation batch was needed and the preregistered test returned KILL. The
substitutions narrowed D1's margin at most from 13 to 12 tasks, and the arm ranking (100/200, 23/59, 14/60) did not
move. Post hoc, reversing D1 would take at least 13 targeted verdict flips among replayed cells and D2's top pair at
least 7 (Appendix~\ref{sec:fidelity}). The margins were known before the seal (13 tasks for D1; 0.110 and 0.156 in
score between adjacent arms for D2), so the test asks whether DGM's recorded endpoints survive the noise re-execution
can expose and says nothing about close decisions. The result is conditional on unreplayed scores: with
submitted tasks and other scores fixed, an unreplayed retained node of a lower-ranked arm would need at least 10 more resolved tasks to tie a higher-ranked arm's best
(initial, unmodified-agent nodes, called roots, lack recorded submission counts and are excluded).
The two decisions are endpoints: the run's intermediate selections are not tested, and one contention profile with
three runs each isolates no causal contention effect.

\begin{figure}[t]
\centering
\includegraphics[width=\columnwidth]{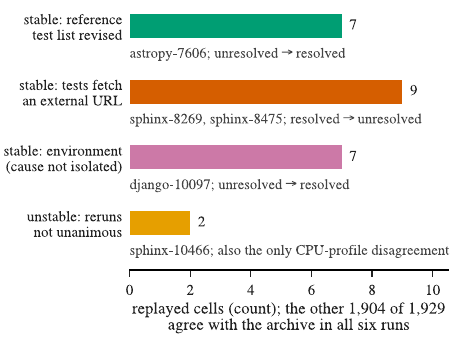}
\caption{Where re-executing DGM's archive disagreed with its recorded verdicts. A cell is one archived evaluation (arm,
agent version, task), re-run three times nominally (four CPUs) and three times under contention (a quarter of a
CPU). Stable drift cells differ from the
record in all six original-host runs. Classifications come from logs and test lists; the Django cause remains
unisolated and its seven drift cells did not reproduce in the second execution. Annotations give the tasks and direction.}
\label{fig:study2}
\end{figure}

\paragraph*{Drift}
The recorded verdict and the rerun majority disagreed on 23 cells (1.19\%) of four tasks, and each disagreement held in
all six runs of its cell (Figure~\ref{fig:study2}). Seven follow a revision of the reference test list that dropped a
required test the archived run had not reported as passing. Nine are on two Sphinx tasks whose required tests fetch an
external URL, which our network-isolated replay cannot reach, whereas the archived runs were networked. Seven, on one
Django task, resolve in every original-host rerun under both the archived and revised test lists; their cause was not isolated. Applying all drift at once still leaves D1 and D2 as recorded.

\paragraph*{Exploratory follow-ups}
Seven original drift cells did not reproduce in a second execution. Rebuilt images changed 16 nominal cell
majorities, eight alongside pytest test-runner and suite-completion differences; causes remain unisolated. SWE-bench 5.0.2
matched all pinned cell verdicts after adapting the archived log format. Reconstructed D1/D2 stayed unchanged
in these comparisons. Suite-incompleteness and test-infrastructure diagnostics use analysis scripts revised after their original seals;
we label them exploratory-unsealed to distinguish them from analyses fixed before the relevant data. Appendix~\ref{sec:c11x} gives scope and deviations. These follow-ups do not recompute
the sealed hypothesis gate.

\section{Implications for grader audits}
\label{sec:lessons}

\paragraph*{Evidence and scope}
Table~\ref{tab:claim-scope} distinguishes decision stability from grader-policy assurance.
Study~2 fixes history and unreplayed scores; testing intermediate selections requires adaptive-search replay.
The extension's two intermittent candidate failures remain unattributed.

\begin{table}[t]
\centering
\footnotesize
\begin{tabular}{@{}>{\raggedright\arraybackslash}p{0.23\columnwidth}>{\raggedright\arraybackslash}p{0.71\columnwidth}@{}}
\toprule
Evidence & Supported scope and boundary \\
\midrule
Study~1 & 0 losses among 59 exposed patches on 25 tasks. Unweighted post-hoc upper bound 4.95\% under candidate independence; 3\% exposed risk unresolved. \\
Study~2 & Best-node selection (D1) and arm ranking (D2) unchanged in the original replay and follow-ups. Original drift partly unreproduced; adaptive learning untested. \\
Extension & 2 rejecting tasks among 59 eligible. Control failed; descriptive only, inconclusive. \\
Selected gold cache & 6 of 15 accepted patches rejected; gold passed on 11 of 11 tasks. Existence only; no prevalence estimate. \\
\bottomrule
\end{tabular}
\caption{Evidence-to-claim summary. Units, exposure and conditioning differ across studies; their tallies cannot be pooled into a common risk estimate.}
\label{tab:claim-scope}
\end{table}

\paragraph*{Lesson 1: plan around policy-exposed units (Study~1)}
For 222 of 281 evaluable patches the arms were identical; only Go patches faced different dependency configurations.
Random task sampling dilutes this contrast. Even all 91 Go tasks, eligible and rejection-free, would leave the
independent-task bound at 3.24\%. At candidate level, 99 independent exposed candidates with zero rejections suffice
for a 3\% upper bound. The extension reproduced 146 candidates, including two with a failing restricted run;
its failed control makes the tally descriptive.
Future audits should sample the policy-exposed stratum directly and check before data collection whether it contains
enough independent units to resolve the tolerance.

\paragraph*{Lesson 2: preregister how intermittent crashes are handled (extension)}
The extension was inconclusive: one gold baseline run had the intermittent crash seen in both candidate rejections,
and every gold run was required to pass. A preregistered stability screen could choose a repetition count $k$ and run the gold patch that many times in the
baseline arm before any policy arm to exclude known-flaky tasks. One training pipeline retries failed gold-patch
runs up to ten times~\citep{auras2026scalingagenticrl}. Such screening changes the conditional population; a later
control failure would still trigger the registered invalidity rule.

\paragraph*{Lesson 3: check replay fidelity before re-executing (Study~2)}
DGM grades filtered patches, and replay images commit setup edits left uncommitted in archived runs.
Size and checksum comparisons caught the patch discrepancy; regrading logs applies no patch and misses it,
though it separated reference-label drift. Matching patch and commit records did not ensure matching verdicts.
Re-executions should check patch and grading inputs and report the runtime environment.

\paragraph*{Lesson 4: validate hermetic graders on accepted alternatives (feasibility stratum)}
The gold patch passed the reconstructed gold-bound cache on all 11 selected tasks, while 6 of 15 accepted alternatives
failed it. Replaying accepted benchmark candidates under a new policy requires container time but no new labels and
detects losses beyond gold-patch validation. Provisioning each candidate's own declared dependencies after
submission recovered every such loss. Validating that this policy also keeps incorrect solutions rejected remains
future work; the confirmatory safety criterion was not met.

\section{Related work}
\label{sec:related}

\paragraph*{Flaky tests and environmental causes}
Network availability~\citep{dietrich2022network}, computational resources~\citep{silva2023raft}
and dependency resolution~\citep{ge2026flakybuilds} affect tests and builds.
We measure consequences for accepted solutions and recorded decisions.

\paragraph*{Reproducible and offline grading}
GitBug-Actions and GitBug-Java capture dependencies so builds re-execute without
network~\citep{saavedra2023gitbugactions,silva2024gitbugjava}, and maintainers apply the idea to grading: SWE-rebench V2
pre-builds task environments~\citep{badertdinov2026swerebenchv2}, Lego-RL packages dependencies into task
images~\citep{du2026legorl} and mounts gold-derived Go module caches~\citep{legorlcode}, and an open, unmerged SWE-bench pull request (as of 2026-10-07) grades
Multimodal tasks offline with bundles bound to the gold patch~\citep{swebenchpr662}.
We test SWE-bench Pro~v1's offline option and a reconstructed Lego-RL gold-bound cache on accepted alternatives.

\paragraph*{Validity of agent benchmarks and rewards}
Inadequate tests admit suspicious patches~\citep{aleithan2024swebenchplus,xue2026swebenchplus}, while test augmentation catches incorrect patches~\citep{yu2025utboost,li2026probe}. A checklist finds outcome-validity violations in
seven of ten agentic benchmarks and OSWorld checkers broken by website changes~\citep{zhu2025abc}. Accepted patches can diverge from
developer intent~\citep{wang2025solved}. OpenAI estimates about 30\% of SWE-Bench Pro's public tasks are broken, including tests rejecting functionally correct submissions~\citep{openai2026signalnoise}. SWE-Bench Pro Verified repairs tasks and audits possible collateral damage in anti-hacking agent runs~\citep{zheng2026sweproverified}; we replay fixed accepted patches under a specified grading policy. Verdicts are compared across
execution substrates~\citep{yuan2026sweminisandbox}, and benchmark ports are validated with parity
experiments~\citep{shi2026harbor}. On rewards, KAT-Coder-V2.5 traces invalid rollouts to sandbox failures~\citep{huang2026katcoder},
outages can corrupt reward signals~\citep{huang2026dsec}, rollout infrastructure carries measurable
overhead~\citep{graviet2026rollouttax}, code RL environments accept incorrect
solutions~\citep{rajan2026rewardhack}, and provable false negatives of math answer verifiers trace to answer
forms~\citep{xin2026verifierfails}. The math-verifier study measures false negatives; we measure code acceptance losses under reference-valid policies.
Test augmentation can overfit a reference implementation~\citep{li2026probe}; Study~1 instead varies dependency availability
with required tests fixed. Study~2 separates reference-label drift.

\paragraph*{Re-execution and self-improvement}
Independent re-implementations of frozen mechanisms audit idea reliability in automated
research~\citep{ning2026lottery}; Study~2 re-executes fixed evaluated patches and separates execution effects from
reference-label drift. Others audit self-improving agents for edits that inflate measured
performance~\citep{wang2026tampering} and argue that coding-agent failures often originate outside the
model~\citep{jarmak2026reliable}.

\section{Conclusion}
\label{sec:conclusion}
The original and follow-up DGM comparisons preserved fixed-history endpoints.
Seven original drift cells did not reproduce on a second host; rebuilt images changed 16 nominal majorities (0.83\%).
The offline policy rejected none of 59 exposed patches, leaving a 3\% risk unresolved; the extension was inconclusive.
Selected cache cases motivate testing accepted alternatives.
Future work should expand exposed samples and replay adaptive search.

\section*{Limitations}
\paragraph*{Coverage and generalization.}
Both preregistered gates returned negative within their tested scope. Study~1 covers one benchmark and published runs
from a few model families at one time. Its policy contrast is Go-only: 59 of 281 evaluable candidates, on 25 tasks,
faced a configuration that the offline policy changes. The motivating 3\% rejection risk remains unresolved.
Study~2 covers one system's archive on one benchmark, with cells from three repositories
(Appendix~\ref{sec:fidelity}). Broader benchmark and archive coverage would allow future studies to test how these
findings generalize, including the effects of intermittent failures and adaptive search. The extension remains
inconclusive because its registered control failed; its tally is descriptive.

\paragraph*{Acceptance-based evaluation.}
Official test acceptance is the operational correctness criterion in this study. Our claims concern acceptance
losses, the quantity a reward or leaderboard consumes; semantic correctness requires separate validation.
Officially unresolved candidates are negatives under this test oracle. The baseline emulates the networked contract
with a file mirror, and candidates needing resources beyond that emulation enter attrition. The reconstructed
gold-bound cache matches the needed module versions in the upstream builder's cache; file identity was not checked.
Semantic review and file-level cache comparisons would extend the evaluation of accepted and unresolved patches.

\paragraph*{Replay controls and selection.}
Study~1 uses three repetitions per arm in fresh containers, baseline reproduction of the official verdict, eligible
controls and a network-unavailability signature. The extension retains the controls but counts any failed restricted
run, a broader rejection rule. Selection requires baseline reproduction and eligibility.
Study~1's 31 unevaluable candidates, including out-of-memory kills, enter attrition rather than the rejection count.
A preregistered baseline stability screen could address intermittent failures in future replays, provided the change
in the eligible population is reported. Replaying excluded cases with larger resource budgets would help characterize attrition.

\paragraph*{Statistical resolution.}
The preregistered gate decisions are the confirmatory claims. Post-hoc bounds and secondary analyses are exploratory;
the Clopper-Pearson bounds assume independent units. A tally below a gate does not supply an upper confidence bound,
and repeated runs do not increase the number of independent tasks. Future sampling plans should target enough
independent, policy-exposed units to resolve the chosen tolerance. Characterizing rare intermittent failures requires
more than the three runs per profile used here. Network-isolated replays fail network-dependent
tests by construction; we classify these differences as drift, separately from instability.

\paragraph*{Prevalence and causal attribution.}
The gold-bound cache rejected 6 of 15 accepted candidates in a deliberately selected feasibility stratum. These cases
establish that the configuration can flip an accepted verdict; prevalence estimation requires a representative sample.
Study~2's original drift count describes one execution, and the second execution reproduced only part of that drift.
Controlled host, image and pytest substitutions would help attribute these differences; the present comparisons
identify neither a host-specific cause nor which verdict is correct. The suite-incompleteness rules are diagnostic
heuristics, and the test-infrastructure intervention leaves edits already present in an image intact
(Appendix~\ref{sec:c11x}). Causal tests of those edits require separate image interventions.

\paragraph*{Version coverage.}
Claims about external systems apply to the tested versions. Appendix~\ref{sec:versions} pairs these pins with upstream
releases checked on 2026-10-07. Study~1 tested SWE-bench Pro~v1 with its v1 evaluator. The maintainers' V2, released
2026-09-22 and now the default configuration, uses a different harness, a networked verifier and new images;
evaluating it is a next step. Study~2 used DGM's pinned 2024 grader commit and September 2025 evaluation images.
Exploratory follow-ups cover a declared log adapter for SWE-bench 5.0.2 and a 2026-10-05 snapshot of rebuilt images,
with reference labels held fixed. Extending coverage to current harness, label and runtime combinations would require
new evaluations beyond the original gate. Appendix~\ref{sec:versions} preserves the dated upstream inventory;
Appendix~\ref{sec:c11x} records the follow-ups.

\section*{Ethical considerations}
This work re-executes publicly released evaluation artefacts: a benchmark's own task images and tests, officially
accepted candidate patches from public leaderboard runs, and one released self-improvement archive. It recruits no
participants and studies no people; the run records contain public identifiers from the
benchmark repositories: commit hashes, commit authors' names and module paths that can include account names. All candidate code and all benchmark tests run
inside network-isolated containers, because the material is third-party code of unknown provenance and the experiment
must not reach the network.

We release verdicts, identifiers and content hashes, not third-party artefacts: candidate patches, task files, reference
solutions, module caches and the released archive are not redistributed, so the replication package neither relicenses
other people's code nor republishes benchmark solutions that could leak into training data. The grading
code we re-execute is released under MIT (the SWE-bench grader and the SWE-bench Pro evaluator); DGM and Lego-RL,
whose patch filter and cache builder we reimplement, are Apache-2.0; the
benchmark datasets' cards state no licence (checked 2026-10-02), and we use them only to evaluate, as intended.

Our findings concern evaluation infrastructure under the conditions reported here. The Limitations section defines
the scope of each gate; broader assurances about agent-evaluation reliability require further evidence.
The appendix reports the studies' compute cost.

\paragraph*{Use of generative AI}
The first author proposed the research direction and directed the study. We used AI assistants for writing,
coding, literature search and internal review. The authors take
responsibility for the methods, results and final manuscript.

\paragraph*{Author statement}
The authors submit this work in their personal capacities as independent researchers.
The views expressed are their own and do not necessarily reflect those of Amazon, Airbnb, Microsoft,
Warner Bros.\ Discovery, Capital One, or any other current or former employer.

\bibliography{references}

\appendix
\section{Replication package}
\label{sec:data}
We provide an anonymized replication package as separate software and data archives in the supplementary material.
A \emph{ledger} lists per-run outcomes; a \emph{manifest} lists input identifiers and content hashes, fingerprints of file bytes. The package contains the
sampling frames, the metadata-only reservation of Study~1's sample, the pinned image digests, the harness, analysis
and figure scripts, both preregistrations of Study~1 and Study~2's preregistration with their dated appendices, the
records of each stage's integrity audit, and the per-run verdict records of every stage (task identifiers, candidate run names,
per-arm outcomes and failure signatures), including the extension's 380 draw records and 3,237 run records. For
Study~2 it contains the archive manifest (node and task identifiers, recorded verdicts, and content hashes of the
evaluated patches, logs and evaluation scripts), the frozen cell sets, the replay and grading scripts, and the verdict
of every run. It also includes anonymized registrations, outcome ledgers, diagnostics and analysis code for the five Study~2 follow-ups
(Appendix~\ref{sec:c11x}); some raw logs and patch hunks remain withheld. The extension's report, the design-limit and negative-control analyses and all figures regenerate from
these committed records (Figure~\ref{fig:arms} also needs FigureSpec, a renderer that turns JavaScript Object Notation (JSON) into vector diagrams, identified in the package's reproduction instructions); the Study~1 and feasibility reports
and the exposed-candidate bound also read raw run directories (the Study~1 report also the evaluator's run scripts), and the
Study~2 report reads the task-subset lists of DGM's code repository, which re-running the experiments or cloning the
pinned repository recreates. Candidate patches, task files, reference solutions,
module caches and the DGM archive are not redistributed; the manifests record identifiers and content hashes for them.
Study~1's task images are pulled by pinned digest, and Study~2 uses locally present images pinned by image
identifier. All candidate code runs only in network-isolated containers.

\section{Compute}
\label{sec:compute}
A container-hour is one container running for one hour; it is not a CPU-core-hour or host elapsed hour.
The original studies execute CPU-only inside network-isolated containers, at most 24 at a time on one host, each pinned to four
logical CPUs with 8\,GiB (Study~2's contention profile caps CPU time at a quarter of one CPU). Study~1's confirmatory run took 93.99 container-hours (5,997 runs; Section~\ref{sec:results}), its
extension 101.44 (the sum over its 3,237 run records), the feasibility stage 80.86 (3,315 runs), and Study~2's replay
68.55 (11,574 runs; Section~\ref{sec:study2}). The task images for Study~1, its extension and the feasibility stage were
pinned by digest: 245 images, 235 of them pulled (about 352 gigabytes (GB) compressed); Study~2 used locally present images pinned by image
identifier. No model was trained or used to assign an experimental test verdict: every such verdict comes from re-executing or
regrading released artefacts. A reviewer model from a different model family ran the design and integrity audits
(Appendix~\ref{sec:sealing}).

\section{Sealing, amendments and audits (Study~1)}
\label{sec:sealing}
The protocol, code, manifests, seeds, sample reservation, analysis script and predictions were sealed by a pushed
commit before any confirmatory data; five amendments are dated appendices committed before the data they affect, and
the runner refuses to start unless the sealed files match. Four post-data appendices correct the reporting and add a
descriptive bound, and change no run, rule or primary quantity. Design and integrity audits were run by a reviewer
model from a different model family in separate sessions; they can make a result inconclusive but cannot pass a gate.
A feasibility stage on separate tasks exercised the harness first; its outcomes did not change the reserved
confirmatory sample.
The connection/name-resolution signature is the case-sensitive \texttt{NET\_RE} pattern in
\texttt{scripts/m0b\_design.py}: \texttt{dial tcp}, Domain Name System (DNS) lookup errors, \texttt{no such host},
\texttt{network is unreachable}, \texttt{Temporary failure in name resolution}, \texttt{ENOTFOUND},
\texttt{EAI\_AGAIN}, or \texttt{getaddrinfo}. The supplementary preregistration
and classifier retain the exact pattern, including the lookup-expression syntax.

\section{The reconstructed gold-bound cache}
\label{sec:cache}
The gold state is the repository after applying the reference patch. Its \texttt{go.mod} declares module dependencies;
\texttt{go.sum} records their checksums. The reconstruction runs the upstream builder's
\texttt{go mod download -json all} and \texttt{go list -deps -test ./...} commands on this state, in the same
pinned image with the same toolchain, with a file proxy and checksum lookups disabled in place of the upstream proxy
and checksum database, seeds the mirror with module metadata files (\texttt{.mod} and \texttt{.info}) for every module version in the gold
\texttt{go.sum} (a registered deviation from the upstream builder; zips only on request), and repeats until they request nothing new. On all 11 tasks the last round of both commands
ended without error output, and no request was a query for a module's latest version. None of the missing module
versions that the six candidates requested is in these caches or in the gold state's \texttt{go.sum}. The checks support
parity of the needed module versions with the upstream builder's cache, not file-for-file parity, because running
that builder requires a networked task container.

The six rejected candidates come from five leaderboard runs of four models: one run lost two tasks and each of the
other four lost one. The stratum was selected for exposure, so these counts describe only how the losses were spread
across runs.

\section{Replay-fidelity checks (Study~2)}
\label{sec:fidelity}
Study~2's 1,929 re-executed cells all come from three repositories: 1,399 on Django, 357 on Sphinx and 173 on
Astropy. Two details affect replay fidelity, and both surfaced before
the seal. DGM grades the patch after removing its diffs to test files and the test-environment configuration file \texttt{tox.ini}, so applying the stored
patch would replay test changes DGM never evaluated; we reimplement the filter, and our evaluated patch matches the
archived one in size and 32-bit cyclic redundancy check (CRC32) checksum for all 6,006 cells. The replay images (SWE-bench's \texttt{v2}-tag evaluation images, distinct from SWE-bench Pro V2 and pinned by image identifier) also commit the harness's setup edits,
which the archived runs left uncommitted, and each run restores the archived layout. Grading uses SWE-bench's grader at
the commit DGM's README pins (\texttt{dc4c087c}, 2024-10-25); regrading all 6,006 archived logs had to agree with the archived verdicts on at least 99\%.

\paragraph*{Unreconstructable staged gate}
Among 89 retained non-root nodes, 38 have fewer recorded successes on the first subset than the released
threshold requires: four successes on the first ten-task subset before expanding evaluation. The archive therefore
cannot reconstruct intermediate selections under that rule.

\paragraph*{Registered substitution and confirmation rules}
For rerun instability (Q1), each comparison uses one matched repetition index across all unstable cells and keeps
every other archived verdict fixed. For contention disagreement (Q3), only cells whose nominal majority agrees with
the archive receive the contention majority. If either substitution reverses an incumbent/challenger pair, its
deciding cells receive three new nominal and three new contention runs, all required to finish. Q1 confirmation
requires the same pair to reverse in at least one matched-index panel using deciding cells that are again unstable.
Q3 confirmation requires the same pair to reverse on cells whose new nominal majority matches the archive and whose
new contention majority repeats the original contention majority. These rules do not count stable drift as support.

\paragraph*{How far the decisions were from reversing (post hoc)}
With unreplayed scores fixed and root entries excluded, reversing D1 needs at least 13 verdict flips among replayed
cells, whether they lower the incumbent or raise the runner-up, reversing D2's top pair at least 7 (the best node without self-improvement gaining 7
of its replayed unresolved cells) and its lower pair at least 10, a tie counting as a reversal in each case. The
registered substitutions changed at most two cells and narrowed D1's margin by at most one task. The 23 drift cells exceed the 7-flip
minimum for D2's top pair, but 22 of them fell on the eight full-system D1 nodes and one on a node without open-ended
exploration, none on the arm without self-improvement, and they cancel on the D1 incumbent and runner-up (net change 0
on each). D2 held because of where drift fell as well as because of its margins. These distances concern reversing the recorded
decisions; they do not show that such flips would occur. This analysis is exploratory and no gate uses it.

\section{The extension in detail (Study~1)}
\label{sec:extension-detail}
This appendix completes Section~\ref{sec:extension}. The extension's unit is the task; unlike the confirmatory run, it requires no
network-unavailability signature and no recovery in A2. The registered tally is two rejection tasks among 59 eligible tasks, one of them the task whose control failed; the other four tasks were
ineligible because the gold patch failed the restricted arm. On both rejection tasks the candidate resolved all
baseline and restored runs and failed one of three restricted runs. Because the control failed, every bound here is
descriptive and the predicted zero was not observed. The one-sided 95\% Clopper-Pearson bounds are 10.29\% on the
share of tasks with a rejection (2 of 59, tasks independent), 4.25\% on the candidate rejection rate (2 of 146
reproduced candidates, candidates independent), and 7.31\% pooled with the prefix's 25 eligible Go tasks (2 of 84 tasks, tasks
independent). Both restricted failures and the failed gold baseline run logged the same nil-pointer panic in Teleport's
\texttt{lib/srv} heartbeat code, with no passing required test. The panic thus also struck the baseline arm, which had
the modules the restricted arm lacks, so it is not confined to the policy; how much the policy contributes is
unmeasured.

\section{Overview of the studies and the Study~1 protocol}
\label{sec:overview}
\begin{figure}[!ht]
\centering
\includegraphics[width=\columnwidth]{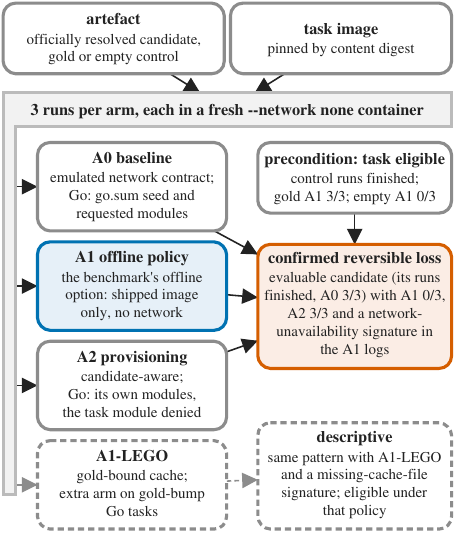}
\caption{Study~1 protocol. Each artefact runs under every arm three times, each run in a fresh network-isolated
container from the pinned image; A0 and A2 add module mirrors for Go tasks only. A confirmed reversible loss counts
only on an eligible task. A1 (A1\textsubscript{BLOCK}) is the offline policy \pblock; Go tasks whose gold patch edits the dependency manifest or checksum file also run under the gold-bound cache
\plego{} (A1\textsubscript{LEGO}, drawn as A1-LEGO, dashed) for a descriptive analysis with eligibility under that policy.
A0 is the baseline and A2 restores candidate-requested modules outside the task's own module path. An eligible task passes the gold patch and fails
the empty patch in all three restricted repetitions. A confirmed loss additionally requires a baseline-reproduced candidate with finished control and candidate cells,
rejection in all restricted runs and acceptance in all restored runs, with the registered network-error signature.
Negative controls are not shown.}
\label{fig:arms}
\end{figure}

Table~\ref{tab:overview} lists the two studies, the extension and the feasibility stratum with their units, contrasts
and outcomes. Figure~\ref{fig:arms} shows how a Study~1 artefact moves through the arms and when a verdict loss counts as
confirmed.

\begin{table}[!ht]
\centering
\footnotesize
\setlength{\tabcolsep}{3pt}
\begin{tabular}{@{}>{\raggedright\arraybackslash}p{0.74in}>{\raggedright\arraybackslash}p{0.78in}
  >{\raggedright\arraybackslash}p{1.33in}@{}}
\toprule
Study and object & Unit (count) & Contrast, event and outcome \\
\midrule
Study~1: SWE-bench Pro, 120-task prefix & accepted patch (281 evaluable; 59 exposed) & emulated networked
  contract vs.\ offline policy; confirmed reversible loss. \textbf{KILL}: 0 losses \\
Extension: remaining 63 Go tasks & task (59 eligible) & same arms; any failed restricted run of a reproduced candidate. \textbf{Inconclusive}:
  control failed; 2 of 59 \\
Study~2: DGM archive & selected evaluation cell (1,929, six runs each) & record vs.\ nominal and contended reruns;
  decision reversal. \textbf{KILL}: 0 reversals \\
Feasibility: 11 selected Go tasks & accepted patch (15) & emulated networked contract vs.\ gold-bound cache. Existence only:
  6 rejected \\
\bottomrule
\end{tabular}
\caption{The studies, their units and their outcomes. Only Study~1 and Study~2 test a preregistered hypothesis with a
kill rule; the extension reports a descriptive tally after its control failure, and the feasibility stratum was selected. Eligible, evaluable and exposed are
defined in Sections~\ref{sec:design} and~\ref{sec:results}.}
\label{tab:overview}
\end{table}

\section{Versions tested and upstream state}
\label{sec:versions}
Every claim about an external system refers to the version we tested. The upstream inventory below is dated 2026-10-07; the five exploratory follow-ups are described separately in
Appendix~\ref{sec:c11x}. Tables~\ref{tab:versions-s1}
and~\ref{tab:versions-s2} give, for each system, the tested pin, the upstream release and the head of its main branch
as checked live and read-only on 2026-10-07 between 15:54 and 16:02 Coordinated Universal Time (UTC), and a class.
\textbf{CURRENT}: the relevant tested component or cited source is unchanged at the checked upstream release or head.
This records component continuity, not a new execution or validation of the latest system as a whole.
\textbf{TESTED-PIN}: the finding refers to a historical pin; newer implementations remain untested.
\textbf{FIXED-UPSTREAM}: the table records a known correction in a later release or commit.
\textbf{UNVERIFIED-CURRENT}: the relevant current configuration was not tested, for the reason given.
Hashes are abbreviated to eight hexadecimal digits; SHA-256 is the file-content hashing algorithm.
A pin fixes a commit, dataset revision or container image. An image ID identifies a locally stored image; a registry
digest identifies an image when it is fetched. Setup edits are changes made while preparing a task environment.
PyPI is the Python package index; cgroups are the host's Linux resource-control mechanism; an out-of-memory (OOM) kill
stops a process that exceeds its memory allocation.

\begin{table*}[!t]
\centering
\footnotesize
\setlength{\tabcolsep}{3pt}
\begin{tabular}{@{}>{\raggedright\arraybackslash}p{0.18\textwidth}>{\raggedright\arraybackslash}p{0.22\textwidth}
  >{\raggedright\arraybackslash}p{0.33\textwidth}>{\raggedright\arraybackslash}p{0.225\textwidth}@{}}
\toprule
System & Tested pin & Upstream on 2026-10-07 & Coverage class and reason \\
\midrule
SWE-bench Pro data (Hugging Face \texttt{ScaleAI/\allowbreak SWE-bench\_Pro})
  & revision \texttt{2d52cb3d}, configuration \texttt{v1}: 731 tasks; parquet SHA-256 \texttt{c8cd7115}, byte-identical to tag
    \texttt{v1.0} (\texttt{307bd006})
  & main is still \texttt{2d52cb3d}; at that revision the \texttt{default} config is V2 (642 tasks, parquet
    \texttt{1067f72c}), with \texttt{hard} and \texttt{v1} as further configs
  & \textbf{CURRENT} for v1: the pin is the head and the v1 config is unchanged \\
\addlinespace
SWE-bench Pro v1 evaluator (\texttt{scaleapi/\allowbreak SWE-bench\_Pro-os})
  & \texttt{swe\_bench\_pro\_eval.py}, run scripts and parsers at \texttt{66f92766}; option
    \texttt{-{}-block\_network} sets network mode \texttt{none}
  & head \texttt{66f92766} is release v2.0.0 (2026-09-22); its V2 README keeps the v1 evaluator and images ``for
    reproducing v1 numbers'' and does not use them for V2
  & \textbf{CURRENT}: the pin is the head and the option is unchanged \\
\addlinespace
SWE-bench Pro V2 (Harbor format)
  & not tested
  & v2.0.0 (2026-09-22), now the default config: 642 tasks, every task's verifier with network mode \texttt{public},
    new images on \texttt{ghcr.io/\allowbreak scaleapi/\allowbreak swe-bench\_pro-v2}; it drops 13 of our 120 prefix
    tasks and 4 of the 63 extension tasks and keeps all 28 prefix Go tasks
  & ``latest protocol keeps network access'' (Section~\ref{sec:background}): \textbf{CURRENT}. Study~1 on V2:
    \textbf{UNVERIFIED-CURRENT}; V2 has no offline verifier, and its images and Harbor verifier were not run \\
\addlinespace
Task images (\texttt{jefzda/\allowbreak sweap-images})
  & by digest: 212 in \texttt{images.json} (182 used in feasibility and the prefix), 63 in \texttt{images\_\allowbreak appendix\_m0c.json}
  & all 212 and all 63 Docker Hub tag digests unchanged; V2 does not use these images
  & \textbf{CURRENT} for v1 \\
\addlinespace
Gold patch fails in every arm (\texttt{flipt-5aef5a14}, \texttt{tutanota-1ff82aa3})
  & v1 data and images
  & V2 drops the Tutanota task; V2's release gate reports the reference patch resolving 642 of 642 tasks, the Flipt
    task included
  & \textbf{UNVERIFIED-CURRENT}: every gold run logs network errors, so replaying without a network may cause it; not re-run on V2 \\
\addlinespace
Offline option fails the gold patch (\texttt{vuls-cc63a0ec}; four extension tasks)
  & v1 evaluator, \texttt{-{}-block\_network}
  & all five tasks are in V2, whose verifier is networked; the v1 option is unchanged
  & \textbf{CURRENT} for the v1 option \\
\addlinespace
Teleport \texttt{lib/srv} panic (\texttt{teleport-10123c04}, \texttt{teleport-a95b3ae0})
  & v1 images
  & both tasks are in V2, on new images
  & \textbf{UNVERIFIED-CURRENT}: not run on V2's images \\
\addlinespace
Lego-RL tooling (\texttt{LegoX/\allowbreak Lego-RL}): gold-bound cache, known-invalid-gold list, quoted cache limit
  & \texttt{ea644fec}
  & head \texttt{a3e28f17} (2026-09-28) is 3 commits ahead, touching four job-configuration and Kubernetes files;
    \texttt{build\_pro\_goproxy.sh} and \texttt{convert\_swebench\_pro.py} are byte-identical; no releases
  & \textbf{CURRENT}: the cache builder and the quoted limit are unchanged \\
\addlinespace
Go toolchain
  & as shipped in each pinned image (exact versions not recorded)
  & go1.27.1 and go1.26.8 (go.dev)
  & \textbf{UNVERIFIED-CURRENT}: newer toolchains not exercised; no toolchain defect is claimed \\
\bottomrule
\end{tabular}
\caption{Study~1: versions tested and upstream state, checked live on 2026-10-07 (15:54 to 16:02 UTC). Classes as
defined in Appendix~\ref{sec:versions}.}
\label{tab:versions-s1}
\end{table*}

\begin{table*}[!t]
\centering
\footnotesize
\setlength{\tabcolsep}{3pt}
\begin{tabular}{@{}>{\raggedright\arraybackslash}p{0.18\textwidth}>{\raggedright\arraybackslash}p{0.22\textwidth}
  >{\raggedright\arraybackslash}p{0.33\textwidth}>{\raggedright\arraybackslash}p{0.225\textwidth}@{}}
\toprule
System & Tested pin & Upstream on 2026-10-07 & Coverage class and reason \\
\midrule
DGM code (\texttt{jennyzzt/dgm}): test-change filter, staged gate
  & \texttt{a565fd2d} (2025-08-13)
  & head is \texttt{a565fd2d}; no releases; no open issue or pull request title concerns the SWE-bench
    evaluation behind D1 and D2 (the nearest, issue \#31 and its open fix \#32, concern parent selection)
  & \textbf{CURRENT}: the pin is the head \\
\addlinespace
DGM archive (\texttt{swe\_results.zip}, linked from DGM's README)
  & Google Drive file of 1,010,489,591 bytes; SHA-256 of its zip listing \texttt{26580b4d}
  & same size, last modified 2025-05-23, same listing hash
  & \textbf{CURRENT}: unchanged \\
\addlinespace
SWE-bench grader
  & \texttt{get\_eval\_report} at \texttt{dc4c087c} (2024-10-25), the commit DGM's README pins
  & PyPI \texttt{swebench} 5.0.2 (2026-08-18); newest tag v5.0.1 (\texttt{87ab1f6c}); main \texttt{02e7a74f}
    (2026-09-02); 33 commits to \texttt{grading.py} and 24 to \texttt{log\_parsers/} since the pin; the 5.0.0
    changelog lists grading fixes
  & \textbf{TESTED-PIN}: Study~2 holds this grader fixed; \textbf{UNVERIFIED-CURRENT} under 5.0.2 and
    main. The grader-version follow-up regrades adapted archived logs with 5.0.2, not current-harness outputs \\
\addlinespace
SWE-bench Verified reference test lists
  & \texttt{SWE-bench/\allowbreak SWE-bench\_Verified} rev \texttt{78f471bf} (parquet SHA-256 \texttt{030cfd7f})
  & head is \texttt{78f471bf} (2026-08-16), same parquet; its commit \texttt{96dae8db} (2026-08-10) dropped
    \texttt{test\_compose\_roundtrip[]} from Astropy~7606's \texttt{PASS\_TO\_PASS} list.
    \texttt{princeton-nlp/\allowbreak SWE-bench\_Verified}, which DGM's code loads, is at \texttt{c104f840}
    (2025-02-18) and keeps it. Issue \#341 on this task was closed on 2025-03-02, cross-referenced by \#338, before
    this change
  & Astropy~7606 label: \textbf{FIXED-UPSTREAM} at \texttt{96dae8db}, in the SWE-bench copy only. Regrade and drift:
    \textbf{TESTED-PIN} grader \\
\addlinespace
Study~2 images (\texttt{swebench/\allowbreak sweb.eval.\allowbreak x86\_64.*})
  & 200 images pinned by image ID, all Docker Hub tag \texttt{v2} (pushed 2025-09-10/11)
  & \texttt{latest} re-pushed for all 200 on 2026-08-13 (163) and 2026-08-16 (37); SWE-bench 5.0.0's changelog is
    dated 2026-08-17; the current recipe (\texttt{SWE-bench/\allowbreak swe-bench-tasks} at \texttt{3d07b464}) still
    commits the setup edits
  & setup-commit detail: \textbf{CURRENT} at the recipe level. Image-dependent results:
    \textbf{UNVERIFIED-CURRENT} at this check; the image-comparison follow-up tests the 2026-10-05 image snapshot \\
\addlinespace
SWE-bench pull request 662 (Section~\ref{sec:related})
  & capture of 2026-09-25: open, unmerged
  & still open and unmerged; one commit, head \texttt{2cdaaa60} (2026-09-18); body unchanged; last updated
    2026-09-28; retitled by its author on 2026-09-18 ``[don't merge, see comment]''
  & \textbf{CURRENT} \\
\addlinespace
Docker Engine (host)
  & 25.0.16 (application programming interface 1.44) on Linux control groups (cgroup v1); contention as a CPU quota (\texttt{-{}-cpus 0.25})
  & \texttt{docker-v29.8.2} (2026-09-30); newest 25.0.x is v25.0.18
  & \textbf{UNVERIFIED-CURRENT}: CPU and memory limits and OOM detection exercised on this engine and cgroup v1 only;
    no engine defect is claimed \\
\addlinespace
Cited system papers (arXiv)
  & versions read: \texttt{2509.16941v2} (SWE-bench Pro), \texttt{2602.23866v2} (SWE-rebench V2),
    \texttt{2609.08149v2} (SWE-Bench Pro Verified), \texttt{2608.17393v1} (Lego-RL), \texttt{2609.22978v1} (DSec),
    \texttt{2505.22954v3} (DGM)
  & each is still the latest arXiv version
  & \textbf{CURRENT} \\
\bottomrule
\end{tabular}
\caption{Study~2, the host and the cited system papers: versions tested and upstream state, checked live on
2026-10-07 (15:54 to 16:02 UTC). Classes as defined in Appendix~\ref{sec:versions}.}
\label{tab:versions-s2}
\end{table*}

\paragraph*{Classification of the findings}
The findings that depend on these components are classified against these versions below; findings that depend on
none of them (for example the arithmetic design limit) need no classification.
These labels describe version coverage; all empirical results retain their tested pins.
\textbf{CURRENT} components underpin: Study~1's KILL, its eligibility counts and the offline option's failure on the reference solution of
the Vuls task and of four extension tasks, because SWE-bench Pro v1, its evaluator and its images are unchanged at the
upstream head; the extension's tally on v1; the feasibility stage's six rejections and the cache limit we quote,
because Lego-RL's cache builder is unchanged; the statement that the maintainers' latest protocol keeps network
access; the staged gate that cannot be reconstructed, the evaluated-patch filter and the recorded arm bests, because
DGM's code is at its head and the archive is unchanged; the setup-commit replay
detail, at the recipe level; the status of the cited pull request; and the cited papers.

\textbf{TESTED-PIN}: the regrade agreement (99.88\%) and the seven Astropy drift cells use the pinned grader and
SWE-bench Verified revision \texttt{78f471bf}. The nine Sphinx network-drift cells were observed when replaying
networked runs without a network; no current-harness replay was performed.

\textbf{FIXED-UPSTREAM}: the empty-parametrization entry in Astropy~7606's reference list (commit
\texttt{96dae8db} of the SWE-bench copy). The princeton-nlp copy that DGM's code loads still has that entry, so which
copy grades decides whether the seven Astropy disagreements arise.

\textbf{UNVERIFIED-CURRENT}, and threats to validity: (i)~Study~1 on SWE-bench Pro V2, which has no offline verifier to
test and needs the Harbor verifier and new images, so its revised task set and environment fixes could change
eligibility and verdicts; (ii)~the Teleport \texttt{lib/srv} panic on V2's images; (iii)~regrade agreement under current-harness logs or main: the grader-version follow-up compares only adapted archived logs under fixed labels;
(iv)~generalization of the original instability bounds and KILL beyond the original execution: the image-comparison follow-up tests a rebuilt-image
snapshot, but does not recompute the sealed hypothesis gate; (v)~behaviour under newer Go toolchains and
Docker Engine releases; (vi)~the two v1 tasks whose gold patch fails in every arm, where every gold run logs network
errors, so the failures may come from replaying without a network (V2 drops one; its networked gate passes the other). The module proxies Study~1 fetched from (goproxy.io and goproxy.cn) are unversioned services;
the Go toolchain checks what they served against \texttt{go.sum} (Section~\ref{sec:design}). These checks were last run
on 2026-10-07.

\section{Exploratory reproducibility follow-ups (Study 2)}
\label{sec:c11x}
The five follow-ups are second-host replay (X1), rebuilt-image comparison (X2), grader-version comparison (X3),
suite-incompleteness diagnostics (X4), and test-infrastructure intervention (X5). The supplementary files use C11X as
their collective label. All five are exploratory; they reconstruct the best-node selection (D1) and arm ranking
(D2) defined in Section~\ref{sec:study2}, without recomputing Study~2's original hypothesis gate.
The second-host comparison was registered while the run was underway but before its verdicts were read. The image
comparison was registered before its runs; the grader comparison followed a disclosed three-cell preliminary check.
The two diagnostic analyses were registered after inspecting earlier follow-up data and use separately revised
scripts. We label those analyses exploratory-unsealed: their final scripts differ from the files fixed by the
original registrations. The original registrations, runners and KILL verdict are unchanged.

\paragraph*{Second-host replay (X1)}
A coauthor repeated all 11,574 runs on a second machine, with the documented host amendment: six four-CPU execution
slots, adjusted CPU affinity (the processors assigned to each run), and scheduler priority. Both executions have
complete outcome ledgers. Verdicts agree at matched run indices on 11,515 runs (99.49\%, descriptive because the
repetitions are independent draws). Nominal and contention majorities each differ on 10 of 1,929 cells.
After validity and coverage checks, the registered reproduction criterion requires at most 1\% nominal-majority
disagreement, retention of every original drift cell, and unchanged reconstructed decisions. A drift cell has a
recorded verdict different from its nominal rerun majority (Q2 in Section~\ref{sec:study2}). The criterion is NOT MET:
seven original drift cells no longer have that disagreement on the second machine, although nominal disagreement
stays below the tolerance and the reconstructed decisions are unchanged.
Those seven django-10097 cells resolve in all 42 original runs and none of the 42 second-machine runs; the latter
contain database errors matching the archived failure signature. Original-host runtime and logs were not inspected
for this comparison, so the cause remains unisolated. Recorded commits, patch hashes and patch-application/evaluation
exit codes match on all runs; this does not establish identical runtime environments. The other differences occurred
on Django cache-expiry/cache-touch tests and the Sphinx task already identified as unstable. The original-host log reading required for
newly gained drift cells could not be completed.

\paragraph*{Rebuilt-image comparison (X2)}
The 2026-10-05 snapshot contains rebuilt container images for all 200 tasks. On the second machine, 5,787 nominal
runs cover all 1,929 cells. Nominal majorities differ from that machine's original pinned-image runs on 16 cells
(0.83\%): eight Astropy, seven Django and one Sphinx. The registered reporting condition is an endpoint change or
more than 1\% nominal-majority disagreement. Neither occurred: reconstructed D1/D2 remain unchanged.
The eight Astropy cells are unresolved in every pinned-image repetition but resolved in every rebuilt-image
repetition. The documented pytest test-runner versions are 7.4.0 and 7.1.2, respectively; pinned-image logs show
suite errors. This observational image comparison does not isolate the effect of substituting pytest. The other
differences involve timing-sensitive and unstable tests. Automatic log classifications check patch-application
failures, timeouts, known reference-label and network-policy cases, and permission-pattern differences; none explains
these changes. Grouped post-hoc diagnostics do not fulfill the registration's requested explanation for each cell.

\paragraph*{Grader-version comparison (X3)}
With reference test lists fixed at dataset revision \texttt{78f471bf}, SWE-bench 5.0.2 without adaptation returns
\texttt{not\_found} on all 6,006 archived logs because they lack its expected test-output delimiters.
The declared adapter adds \texttt{>>>>> Start Test Output} and \texttt{>>>>> End Test Output} around each archived log,
preserving the original log content, changing no grader code and supplying no test exit code. The adapted logs receive the same
resolved/unresolved verdict as under the pinned grader on every cell. This comparison treats 709 adapted
\texttt{not\_found} outcomes as unresolved because the adapted logs contain no parsed test results: \texttt{not\_found} is the grader's status for output it does not accept as
usable test evidence, rather than a third experimental verdict. Seven per-test success/failure assignments differ
on cells unresolved under both graders. Thus cell-verdict agreement does not establish equivalence of the parsers
or grading rules. The best node and arm ranking are unchanged, although the score gaps between challengers and the
best node differ from the reconstruction using recorded verdicts.
This comparison reconstructs all 96 nodes with run reports, a larger set than the final-archive node sets fixed for
the original study. The three initial, unmodified-agent nodes lack recorded submission counts; their task-count
denominators are recovered from their fixed scores. The comparison neither re-executes tests nor covers logs
produced by the current harness or revised reference test lists.

\paragraph*{Suite-incompleteness diagnostics (X4)}
The scan flags a run if its log meets any of three checks: the pinned grader's parser finds no test results; the
final pytest summary reports errors and has at least as many errors as passed tests and no required \texttt{FAIL\_TO\_PASS} test has a
passed/failed status; or a traceback occurs after the last result line recognized by the scan's pattern (or no
result line is recognized). A cell is flagged only when all three nominal runs are flagged. The stricter variant
retains the empty-parser and error-summary checks but removes the traceback check.
A revised script flags 189 of 1,929 pinned-image cells (9.8\%) under the original checks and 150 (7.8\%) under the
stricter variant. Thirteen cells are flagged on pinned images but not rebuilt images; eight become resolved.
These are log-based flags, not a causal classification of environment failures. The traceback check also flags
completed suites, including 15 archived cells recorded as resolved. Neither variant captures the known django-10097
failure family. Both meet the registered reporting threshold: at least 1\% of cells, with at least two exception
signature families each covering at least three cells. A run's family is its most frequent match from a fixed list
of exception/warning names, or \texttt{other}; a cell takes its most common run label. These are text-pattern groups. Meeting this descriptive threshold establishes neither
causal attribution nor a new defect.

\paragraph*{Test-infrastructure intervention (X5)}
The scan flags files by basename: \texttt{conftest.py}, \texttt{pytest.ini}, \texttt{setup.cfg},
\texttt{pyproject.toml}, \texttt{setup.py}, \texttt{.coveragerc}, \texttt{sitecustomize.py},
\texttt{usercustomize.py}, and \texttt{eval.sh}, plus files ending in \texttt{.pth} (Python path-configuration files).
Among 6,006 archive cells, 1,291 have an evaluated patch that touches one of these files. In 1,203 of those cells,
the only flagged edits match dependency-version pins inserted by the harness's pre-install commands. This
classification uses strings in the added lines, rather than establishing that the agent authored those edits.
A revised scan selects 59 of the 769 recorded-resolved replay cells (7.67\%) for one run each after removing the flagged files' diff
blocks, meaning all patch changes for those files. Fifty-three resolve and six do not. The registered rule labels
the result SUPPORTED when at least 1\% of recorded-resolved replay cells touch flagged files and at least one
complete-suite run becomes unresolved after block removal, compared with its archived resolved verdict.
All six failures also occurred in full-patch nominal runs on both machines. Comparing against archived acceptance
therefore gives no new evidence that removing a block caused the failure. Block removal also leaves environment
edits already present in an image intact. This selected, one-run intervention cannot establish the absence of reward
manipulation in the archive.

\paragraph*{Evidence and resources}
The supplementary package contains the outcome ledgers and diagnostics used for these counts, hashes recorded when
the original protocols and analysis files were fixed, and a manifest identifying changes made to anonymized export
copies. Raw logs and some patch-hunk records are withheld; the export cannot regenerate every source-level
diagnostic. Recorded run durations sum to 50.51 container-hours for second-host replay and 12.59 for the image
comparison. The infrastructure intervention adds 59 checks whose durations are not exported. The grader comparison
uses one container to grade archived logs, and the suite diagnostics scan retained logs. No model was trained or
assigned an experimental verdict.

Internal reviews checked scoped follow-up analyses, claims, statistics and rejection arguments.
A citation-focused check covered five added references and nine other citation contexts. These checks do not
constitute an independent reproduction of the experiments or verification of every final file.
A separate reproduction check regenerated ledger summaries and, with pinned dependencies and fonts, source
scalable vector graphics (SVG) figures; its scope did not include re-executing experiments or certifying the full
accompanying package. Font substitution can change regenerated PDF figures even when source SVGs match.

\end{document}